 \documentclass[twocolumn]{aastex631}

\shorttitle{Optical Rise of V1674 Her}
\shortauthors{Hachisu \& Kato}

\begin{document}

\title{Optical detection of the X-ray flash in the very fast nova V1674 Her:\\ 
Optical contribution of the irradiated accretion disk}


\author[0000-0002-0884-7404]{Izumi Hachisu}
\affil{Department of Earth Science and Astronomy,
College of Arts and Sciences, The University of Tokyo,
3-8-1 Komaba, Meguro-ku, Tokyo 153-8902, Japan}
\email{izumi.hachisu@outlook.jp}

\author[0000-0002-8522-8033]{Mariko Kato}
\affil{Department of Astronomy, Keio University,
Hiyoshi, Kouhoku-ku, Yokohama 223-8521, Japan}




.


\begin{abstract}
V1674 Her is one of the fastest and brightest novae, characterized by
dense optical photometry in the pre-maximum phase, a rise from $g=17$
to 7 mag, in one-fourth of a day.
We present a composite theoretical $V$
light curve model of its early rising phase starting from a quiescent
brightness of $g=19.2$ mag.  Our light curve model consists of a hot
and bright white dwarf (WD) and irradiated accretion disk and companion
star.  We found that the earliest optical detection of ASAS-SN $g$ band
brightness of $g=17.0$ at $t=0.014$ day from the onset of thermonuclear
runaway can be explained with the irradiated accretion disk and companion
star in the X-ray flash phase of a $1.35 ~M_\sun$ WD.  
This is the first detection in optical of an X-ray flash phase of a nova.
Optically thick winds emerge from the WD photosphere at $t=0.04$ day,
and optical flux is dominated by free-free emission from optically-thin
ejecta just outside the WD photosphere.
Our free-free emission model $V$ light curve reasonably reproduces the dense
$g$ light curve of Evryscope that spans from $g=14.8$ (at 0.078 day)
to $g=7.1$ (at 0.279 day), including a sudden change of slope in 
the $g$ light curve from slow to rapid rise at $g=14.3$ on day $0.1$.
There is no indication of shocking power during the rising phase from 
$g=14.8$ to 7.1.
\end{abstract}






\section{Introduction}
\label{introduction}


\begin{figure*}
\epsscale{0.9}
\plotone{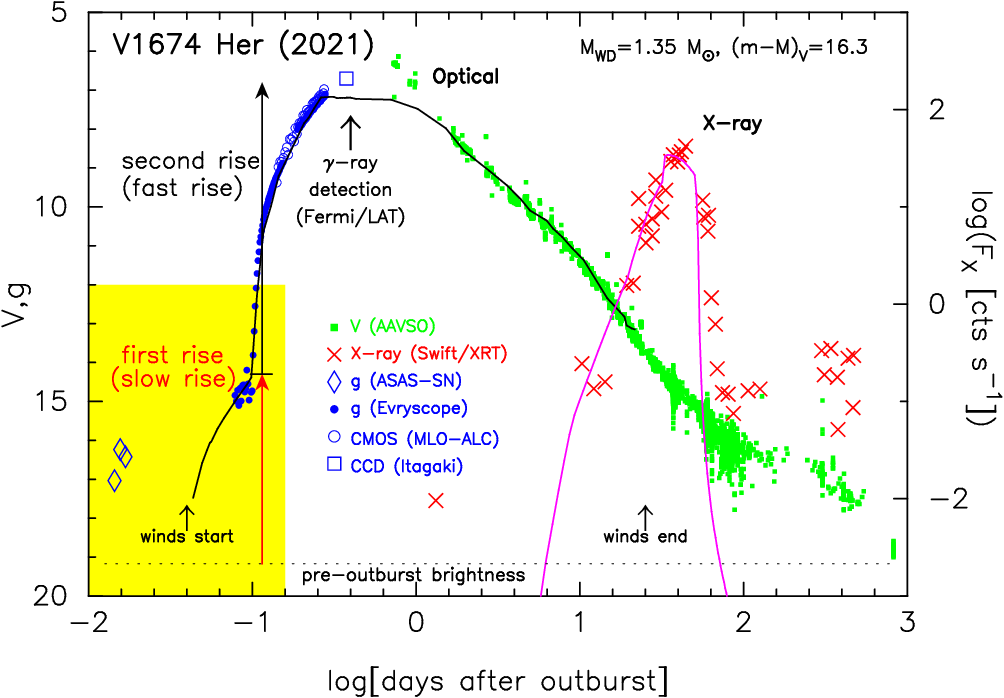}
\caption{
The optical $V/g$ and X-ray (0.3--10.0 keV) light curves of V1674 Her.
The $V$ data are taken from the archive of 
the American Association of Variable Star Observers (AAVSO).
The All-Sky Automated Survey for Supernovae (ASAS-SN) $g$, 
Evryscope $g$, and Itagaki's unfiltered CCD data are from \citet{qui24}.
The X-ray count rates are from the Swift website \citep{eva09}. 
The Mount Laguna Observatory All-Sky Camera (MLO-ASC) data are
from \citet{sok23}.  We add a theoretical $V$ and X-ray light curve
based on \citet{kat25hs}'s fully self-consistent nova outburst model.
We set our theoretical outburst day (stage B in Figure \ref{hr})
to be HJD 2,459,377.68. 
The white dwarf (WD) model has the mass of
$M_{\rm WD}= 1.35 ~M_\sun$ with the mass-accretion rate of
$\dot{M}_{\rm acc}= 1\times 10^{-11} ~M_\sun$ yr$^{-1}$.
The model $V$ light curve (black line) is calculated from free-free emission
from nova winds whereas the model X-ray light curve (magenta line) is
calculated from the blackbody emission from the photosphere (0.3--10.0 keV).
The $V$ band distance modulus $\mu_V\equiv (m-M)_V= 16.3$,
the distance $d=8.9$ kpc, and the extinction
$E(B-V)=0.5$ are taken from \citet{kat25hs}. 
There is a sudden slope change (a break of the black line)
in the rising phase, as demonstrated in the yellow-shadowed area.
We divide the rising phase into two, before and after the break
as shown by the red and black upward arrows.
We also show the quiescent brightness of $g = 19.17$ \citep[dotted 
line; ][]{qui24} 1.7 days before the nova outburst ($t=-1.7$ day).
\label{v1674_her_v_x_observation_only_logscale}}
\end{figure*}

A nova is a thermonuclear runaway event on a mass accreting white dwarf (WD).
After hydrogen ignites at the bottom of the WD envelope,
the WD brightens up and its hydrogen-rich envelope expands largely to 
emit strong winds \citep[e.g., ][for a recent fully self-consistent 
nova outburst model]{kat22sha}.
Many classical novae have ever been detected only after or close to their 
optical maxima.  Therefore, their very early pre-maximum phases have not
been studied in detail.

  One of the rare exceptions is the X-ray flash
detection in the classical nova YZ Ret \citep{kon22wa}. 
An X-ray flash is a soft X-ray bright event of a nova
just after the thermonuclear runaway starts, typically 
a few days before optical maximum in very massive WDs 
like in YZ Ret \citep{kat22shapjl, kat22shc}.  
From an X-ray spectrum analysis of YZ Ret, \citet{kon22wa} found no major
intrinsic absorption during the X-ray flash.
Comparing the YZ Ret X-ray data with their model calculations,
\citet{kat22shapjl} concluded that, in such an early phase of a nova outburst,
\\
(1) no dense matter exists around the WD photosphere,\\ 
(2) no indication of a shock wave, and\\
(3) the hydrogen-rich envelope is almost hydrostatic.\\
These conclusions are very consistent with their theoretical models
in which massive optically thick winds emerge from the WD photosphere
only after the X-ray flash phase ended. 

Another rare exception is 
the very fast nova V1674 Her (Nova Herculis 2021) discovered 
on UT 2021 June 12.537 by Seiji Ueda (cf. CBET 4976). 
This object has been observed in multiwavelength,
from radio, optical, UV, X-ray, to gamma-ray
\citep[e.g.,][]{dra21, woo21, lin22, pat22, ori22, sok23, bha24,
hab24, qui24}.  Early observations are summarized 
in \citet{dra21}, \citet{sok23}, \citet{bha24}, and \citet{qui24}. 




A remarkable feature is dense optical photometric data in the rising phase,
which is shown in Figure \ref{v1674_her_v_x_observation_only_logscale}. 
The source of each data can be found in the caption. 
A plenty of optical data were obtained from $g=17.0$ to $g=7$,
over 10 magnitudes rise only during one-fourth of a day
\citep{sok23, qui24}.


\begin{figure*}
\epsscale{0.85}
\plotone{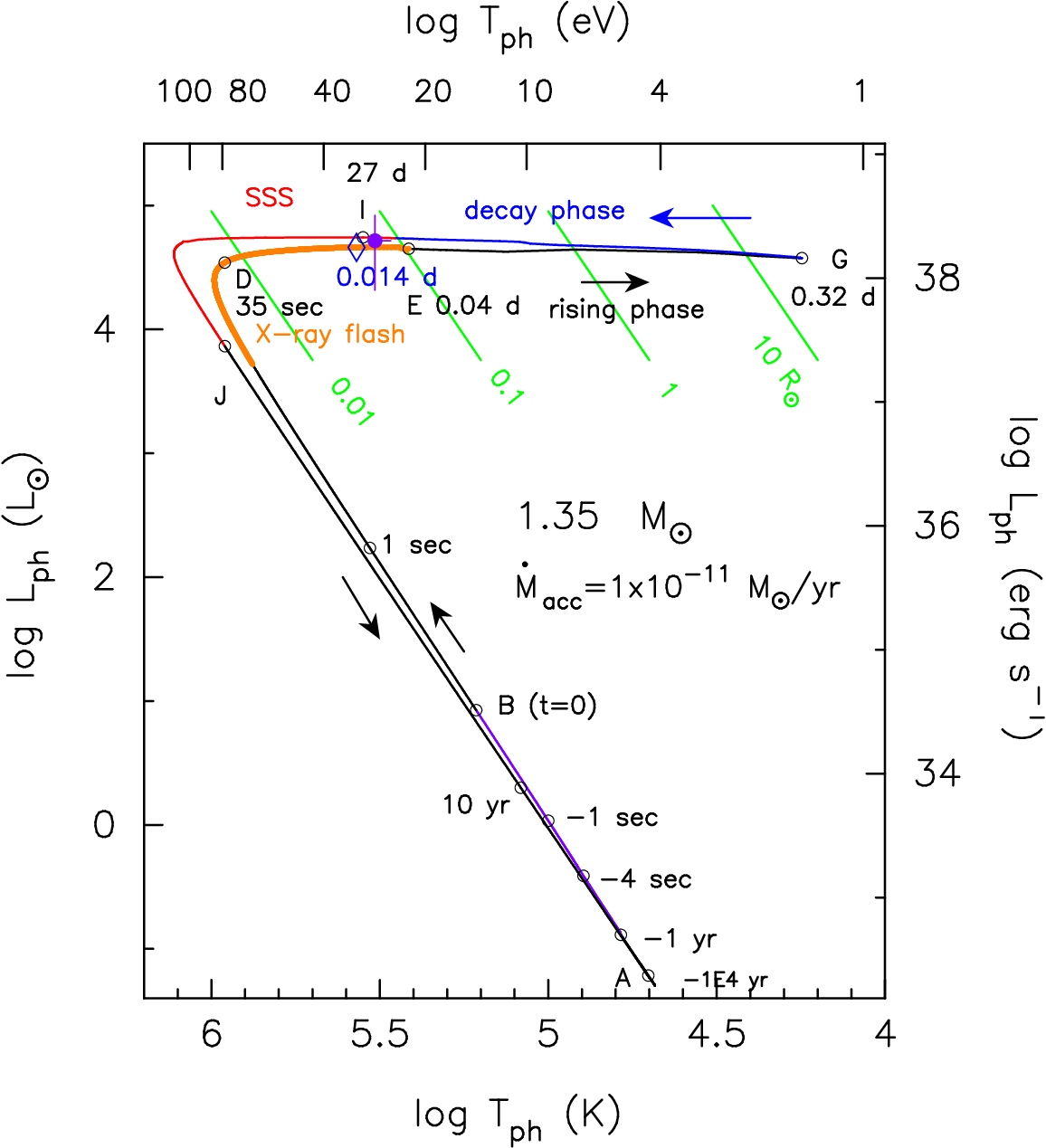}
\caption{
The H-R diagram of one cycle of hydrogen shell flashes for
our nova outburst model of a $1.35 ~M_\sun$ WD with the mass 
accretion rate of $\dot{M}_{\rm acc}=1\times 10^{-11} ~M_\sun$ yr$^{-1}$,
taken from Figure 1(a) of \citet{kat25hs} with several modifications.
Only the photospheric luminosity $L_{\rm ph}$ and temperature $T_{\rm ph}$
of the exploding WD are plotted.  Note that the contributions from 
those outside the WD photosphere, e.g., the WD winds (ejecta),
irradiated disk, and companion star, are not included.
Selected stages during a shell flash are denoted counterclockwise direction 
starting from the bottom of the cycle.  
A: quiescent phase before the shell flash. 
B: the epoch when $L_{\rm nuc}$ reaches maximum ($t=0$).
D: the peak of soft X-ray luminosity $L_{\rm X}$ (0.3--1.0 keV). 
E: winds emerge from the photosphere. 
G: the maximum expansion of the photosphere and maximum wind mass loss rate.
The straight green lines depict the locus of constant photospheric
radius of $R_{\rm ph}= 10$, 1, 0.1, and $0.01 ~R_\sun$. 
The times at selected stages are indicated before and after stage B.
The open blue diamond labeled ``0.014 d'' indicates the epoch of
the first ASAS-SN detection of the V1674 Her outburst in Figure 
\ref{v1674_her_v_x_observation_only_logscale}. 
The filled purple circle with error bars indicates the detection 
of YZ Ret in the X-ray flash phase \citep{kon22wa}. 
See Figure 1(a) of \citet{kat25hs} for more details and other symbols. 
\label{hr}}
\end{figure*}

This plot also shows a theoretical 
free-free emission model $V$ light curve 
(black line) and X-ray (0.3--10 keV) light curve in the 
supersoft X-ray source (SSS) phase (magenta line), the data of which are
taken from \citet{kat25hs} for a $1.35 ~M_\sun$ WD with a mass 
accretion rate of $1 \times 10^{-11}M_\sun$ yr$^{-1}$. 
Their model $V$ light curve reproduces overall properties of
the observed data, from the very fast rising phase of $V \sim 14$ mag 
until the end of the outburst. 
This is the first successful model that reproduces the rapid rise
of a classical nova in the very early phase.  

\citet{kat25hs} did not show the earliest rising phase
from the quiescent brightness of $g=19.17$ to $g\sim 14$ mag.  
The ASAS-SN $g$ data (open blue diamonds) seems to be in a different stream
from the Evryscope data (blue dots).  
The aim of this paper is to elucidate  
this very early phase of the outburst of V1674 Her 
with a binary model including irradiation effect by the hot and bright WD.
Our $V$ light curve model consists of the exploding WD (taken from
\citet{kat25hs}'s $1.35 ~M_\sun$ WD model), disk and companion star
both irradiated by the hot and bright WD.  

This paper is organized as follows. First we explain the one cycle of
the nova outburst in the H-R diagram based on the $1.35~M_\sun$ WD model
\citep{kat25hs} in Section \ref{sec_model}.
Then, we introduce our binary model that consists of a hot and bright WD, 
an irradiated accretion disk and companion star in Section
\ref{sec_binary_model}.
We compare our model light curves with observational results of V1674 Her
in Section \ref{first_x-ray_flash}.
Discussion and conclusions follow in Sections \ref{sec_discussion}
and \ref{sec_conclusion}, respectively.


\section{Basic nova model}
\label{sec_model}

\subsection{H-R Diagram of one cycle of the nova outburst}

Figure \ref{hr} shows one cycle in the H-R diagram of the nova outburst 
calculated by \citet{kat25hs}, i.e., a $1.35 ~M_\sun$ WD with a mass
accretion rate of $1\times 10^{-11} ~M_\sun$ yr$^{-1}$.  
This plot is essentially the same as Figure 1(a) of \citet{kat25hs}
with several modifications.   
In the quiescent phase (inter-outburst period), the accreting WD stays
around the bottom of the loop. 
After a thermonuclear runaway starts, the WD moves quickly upward 
keeping the photospheric radius almost constant. 
We define the origin of time (stage B, $t=0$) by the epoch 
when the nuclear energy generation rate 
reaches maximum, $L_{\rm nuc}=L^{\rm max}_{\rm nuc}$.  
After that, the photospheric temperature continuously
increases to its maximum at $\log T_{\rm ph}^{\rm max}$ (K)= 6.02.  
With such a high temperature, the WD photosphere emits soft X-ray  
photons, which is called the X-ray flash phase.
Then, the WD envelope expands and the photospheric temperature
turns to decrease. 

The epoch of the first optical detection of the V1674 Her outburst
\citep[ASAS-SN $g=17.0$ on day 0.014, ][]{qui24} is plotted
by the open blue diamond labeled ``0.014 d'' on the H-R track.  

Optically thick winds emerge from the WD photosphere 
when the envelope expands and the photospheric
temperature decreases to $\log T_{\rm ph}$ (K)=5.41
(stage E, the open circle labeled E).
Soft X-rays could be absorbed by wind itself far outside the photosphere,
thus, we regard this to be the end epoch of the X-ray flash phase.  

We showed the part of X-ray flash phase with the orange line
on the track, that is, from the epoch of $L_{\rm X}=10^{3.2} ~L_\sun$
($=0.1 L_{\rm X}^{\rm max}$) to stage E. 

The photosphere further expands and reaches maximum at stage G.
The wind mass loss rate also increases and reaches maximum at the same
time at stage G.
The hydrogen-rich envelope quickly loses its mass 
mainly due to wind mass loss.
After stage G, the photosphere turns to shrink with time 
while the photospheric temperature turns to increase.
The optically thick winds stop
at epoch I when the photospheric temperature
increases to $\log T_{\rm ph}$ (K) =5.55.
The WD emits soft X-ray photons (the supersoft X-ray source (SSS) phase).
As hydrogen burning decays, the WD becomes faint and returns to 
the quiescent phase (the bottom of the loop). 

In the present work, we focus on a very early phase of 
the outburst, i.e., from stage B ($t=0$) to the
X-ray flash phase (orange line part on the H-R track),
and to the ensuing rising phase (black line part) of the WD.

\subsection{Free-free emission model $V$ light curve}
\label{free-free_emission_light_curve}

After the optically thick winds emerge from the photosphere,
we have calculated free-free emission luminosity from the optically thin
ejecta outside the photosphere \citep[e.g.,][]{hac06kb, hac20skhs}.
While optically thick winds are
accelerated deep inside the photosphere, the nova wind itself becomes
optically thin outside the photosphere.
The $V$ luminosity can be simplified as
\begin{equation}
L_{V, \rm ff,wind} = A_{\rm ff} ~{{\dot M^2_{\rm wind}}
\over{v^2_{\rm ph} R_{\rm ph}}},
\label{free-free_flux_v-band}
\end{equation}
where $\dot{M}_{\rm wind}$, $v_{\rm ph}$, and $R_{\rm ph}$ are
the wind mass loss rate, wind velocity at the photosphere, 
and photospheric radius, respectively, calculated by \citet{kat25hs}.
See Equation (3) in \citet{kat25hs} for details on the coefficient
$A_{\rm ff}$ and how to determine it for V1674 Her.

We depict the calculated $L_{V, \rm ff,wind}$ (free-free emission
luminosity) by the black line in Figure 
\ref{v1674_her_v_x_observation_only_logscale}
based on \citet{kat25hs}'s nova evolution calculation.
It is remarkable that only the free-free emission model $V$ light curve
reasonably reproduces the entire nova $g/V$ light curve in the wind
phase, from a very early phase of $g=15$ mag (at 0.078 day)
until $V=13.3$ mag (on day 22.4). Here, we assume that $g-V=0$ \citep{qui24},
and adopt the $V$ band distance modulus toward V1674 Her of
$\mu_V\equiv (m-M)_V= 16.3$, distance of $d=8.9$ kpc, and extinction of
$E(B-V)=0.5$ after \citet{kat25hs}.

The wind phase begins at 0.04 day and ends on day 26.
Almost throughout this interval, the free-free
emission model $V$ light curve reasonably reproduces the observed
$g/V$ light curve from $g=15.0$ to $V\sim 13$.
This indicates that the optical contribution from each photosphere of
the disk, companion star, or WD is negligible from $g=15.0$ (at 0.078 day)
to $V= 13.3$ (on day 22.4).


\begin{figure*}
\gridline{\fig{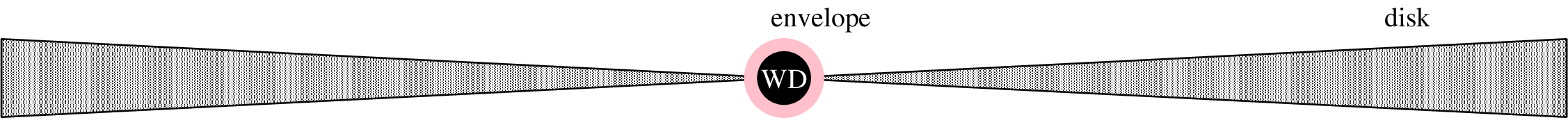}{0.75\textwidth}{(a) the expanding envelope just 
reaches the inner edge of the disk}
          }
\gridline{\fig{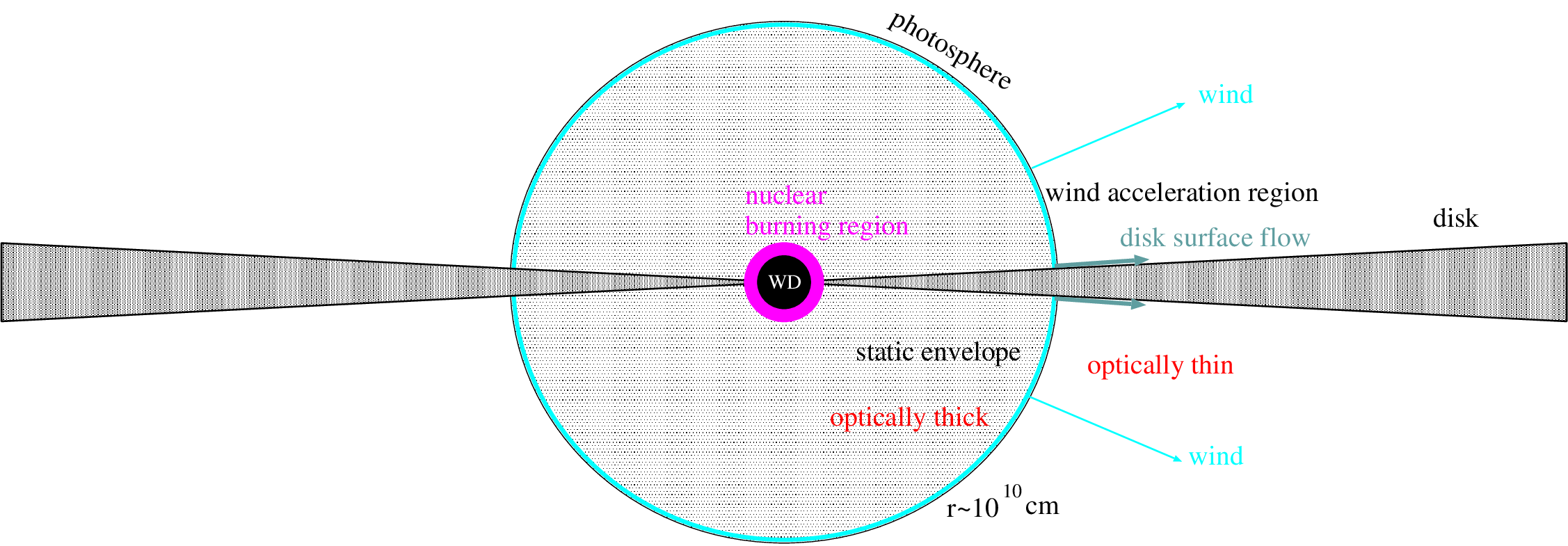}{0.75\textwidth}{(b) winds begin
to emerge from the photosphere at epoch E}
          }
\gridline{\fig{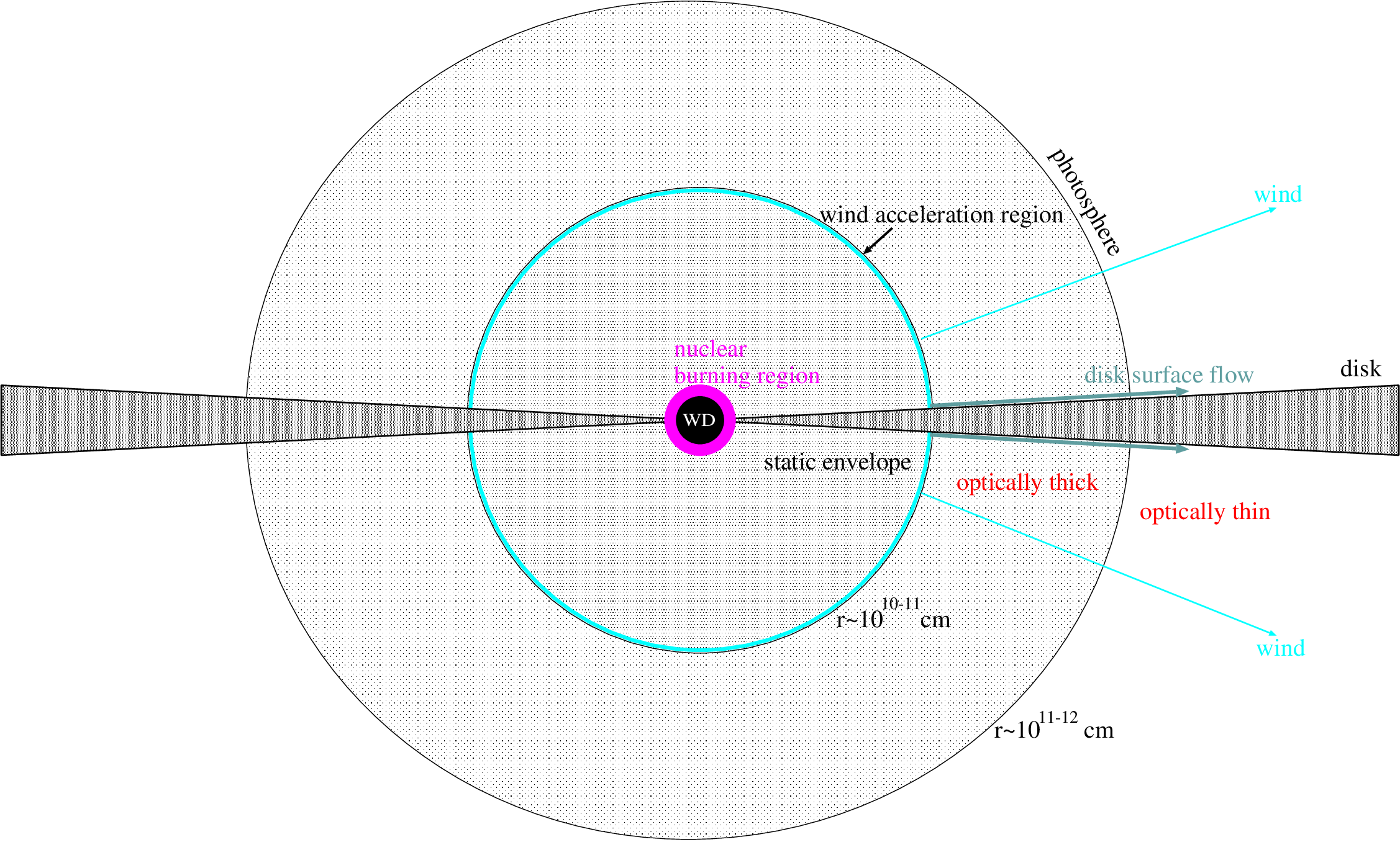}{0.75\textwidth}{(c) nova wind phase}
          }
\caption{
Schematic configurations of a WD envelope and accretion disk
during a nova outburst.
(a) In the X-ray flash phase: when the expanding envelope reaches the 
inner edge of the accretion disk. 
(b) At epoch E: The optically thick winds emerge from the photosphere. 
The accretion disk is not disrupted and the winds blows avoiding the region 
of the disk.  
(c) Early wind phase after stage E: The disk has not 
been totally engulfed yet by the WD envelope. The critical point of
wind acceleration is at $R_{\rm cr}\approx 0.2 ~R_\sun$. 
Optically thick winds have an opening angle avoiding the accretion disk. 
The disk surface flow occurs owing to the Kelvin-Helmholtz instability or
friction between the winds and the disk surface. 
The spherical nova envelope configurations are taken from the spherically
symmetric $1.35 ~M_\sun$ WD model \citep{kat25hs}.
The thick light-cyan region describes the acceleration region of winds.
The dark-cyan arrows represent the disk surface (boundary layer) flows,
which exist near the surface of accretion disk
and have slower outflow velocities than those of winds.
Here, we neglect the inner magnetic polar accretion region/stream because
they are embedded by the almost static hydrogen-rich envelope rather inside
the acceleration region.
\label{v1674_her_disk_config}}
\end{figure*}

\subsection{X-ray flash phase}
\label{x-ray_flash_phase}

As introduced in Section \ref{introduction}, the first
detection data of ASAS-SN (open blue diamonds at $t=0.014$ day in 
Figure \ref{v1674_her_v_x_observation_only_logscale}) was obtained
much earlier than when our free-free emission model $V$ light curve
(black line) rises up.  This epoch ($t=0.014$ day) corresponds to
the X-ray flash phase, as shown by the open blue diamond in the H-R
diagram (Figure \ref{hr}).
It is interesting that this position is very close to that of YZ Ret
during its X-ray flash phase in the H-R diagram, of which the temperature
and luminosity are estimated by \citet{kon22wa}. 
YZ Ret is a very fast nova of which the WD mass is estimated
to be very massive  \citep[$>1.3~M_\sun$, ][]{kat22shapjl, kat22shc}.
This resemblance also suggests that the ASAS-SN $g=17.0$
data was obtained during its X-ray flash phase.
In the X-ray flash phase, the hydrogen-rich envelope of the WD 
is too hot ($T_{\rm ph}\sim 3\times 10^5$ K) to be bright in optical.
This leads us to include the brightnesses of 
an accretion disk and companion star irradiated by the hot and bright WD.
In what follows, we will see whether or not the irradiated disk and
companion star are enough to reproduce these $g$ band brightnesses
of $g=$17.0--16.2 on day 0.014--0.017.

\section{Binary model}
\label{sec_binary_model}

\subsection{Binary parameters}

Our binary model consists of a $1.35 ~M_\sun$ WD \citep{kat25hs},
accretion disk, and companion star.
Both accretion disk and companion star are 
irradiated by the hot and bright WD.
We adopt a companion mass of $M_2= 0.26 ~M_\sun$ after \citet{qui24},
the orbital period of 0.152921 day ($=3.67$ hr) and ephemeris of
Minimum Light $=$ HJD $2,459,400.637 ~+ ~0.152921 ~E$
after \citet{pat22}.  We assume that the epoch of minimum light
corresponds to the configuration that the companion star is just in front
of the WD.  Then, we obtain the separation of
$a= 1.41 ~R_\sun$, the effective Roche lobe radii of 
$R_{\rm RL,1}= 0.739 ~R_\sun$ and 
$R_{\rm RL,2}= 0.351 ~R_\sun$, and the orbital velocities of
$v_1 = 75.4$ km s$^{-1}$ and $v_2 = 391$ km s$^{-1}$.
The inclination angle of the binary/accretion disk
is assumed to be $i=67\arcdeg$ after \citet{hab24}.
Among above parameters,
only the inclination angle is not well constrained, so we calculated
with other two inclination angles of $i=45\arcdeg$ and $i=75\arcdeg$ to
examine the brightness variation.



\begin{figure}
\epsscale{1.15}
\plotone{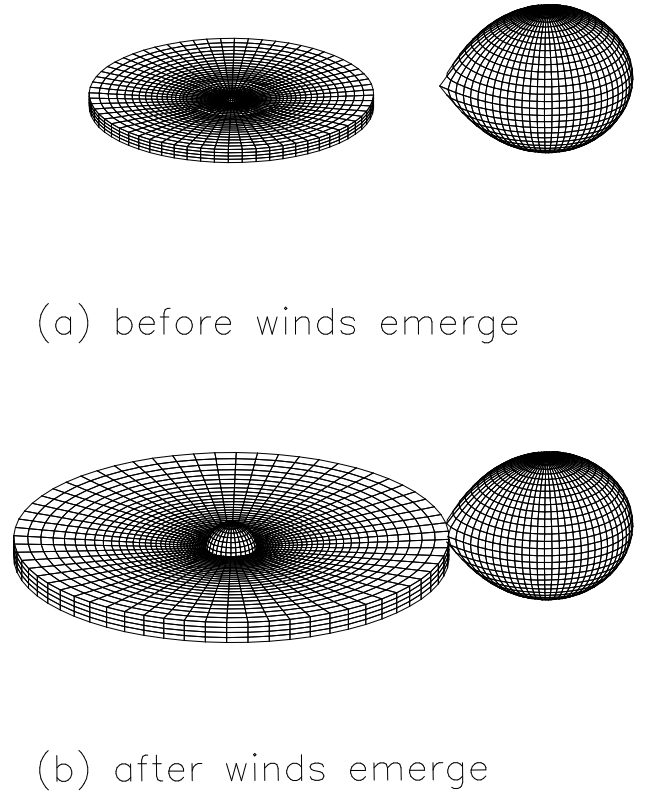}
\caption{
Geometric configuration models of our disk and companion star
in Figure \ref{comparing_light_curve_v1674_her_early}(a).
The masses of the WD and Roche-lobe-filling companion star
are $1.35 ~M_\sun$ and $0.26 ~M_\sun$, respectively.
The orbital period is $P_{\rm orb}= 0.1529$ day ($=3.67$ hr).
The inclination angle of the binary is $i=67\arcdeg$.
The separation is $a= 1.41 ~R_\sun$ while their effective
Roche lobe radii are $R_{\rm RL, 1}= 0.739 ~R_\sun$ and
$R_{\rm RL, 2}= 0.351 ~R_\sun$.  We assume, in panel (a), the disk size
is $R_{\rm disk}= 0.628 ~R_\sun$ ($=0.85~R_{\rm RL, 1}$), the height of
the disk edge to be 0.05 times the disk size but, in panel (b), the disk
size is $R_{\rm disk}=0.961 ~R_\sun$ ($=1.3~R_{\rm RL, 1}$)
and the edge height is 0.05 times the disk size.
The photospheric surfaces of the disk and companion star are irradiated
by the central hot WD. Such irradiation effects are all included in the
calculation of the $V$ light curve reproduction \citep[see][for the
partition of each surface and calculation method of irradiation]{hac01kb}.
We also include the effect of viscous heating in the accretion disk
for a given mass-accretion rate \citep{hac01kb}.
We neglect the inner edge of the disk mainly because the optical
brightness is determined by the surface area of the disk and 
a central hole (very small area) hardly contributes to the optical brightness.
\label{v1674_her_config}}
\end{figure}

\subsection{Disk is not disrupted}
\label{not_disrupted_disk}

Theoretically, we expect that the accretion disk remains undisrupted
during the nova outburst. 
As already explained in Section \ref{introduction},
the WD envelope expands 
almost hydrostatically during the X-ray flash phase.
In our nova model, the photosphere of expanding envelope reaches
the inner edge of the accretion disk,
e.g., at $t=5.3$ min ($=0.0036$ day).
Here, we assumed the inner edge of the disk to be 
$R_{\rm in}\approx 0.02 ~R_\sun$ after \citet{qui24}
for the intermediate polar (IP) V1674 Her.
The disk is almost intact because optically thick winds
do not yet emerge from the photosphere, as illustrated in Figure
\ref{v1674_her_disk_config}(a).

After that, the WD photosphere continues to expand keeping a hydrostatic
balance. The inner disk is engulfed by the envelope, but not disrupted
because the envelope is in hydrostatic equilibrium
and its density is much smaller
than that of the accretion disk. The optically thick winds begin
to blow from the WD photosphere ($R_{\rm ph}
\approx 0.1 ~R_\sun$) on day 0.04 (stage E).
The acceleration region appears close to the photosphere
as shown in Figure \ref{v1674_her_disk_config}(b)(thick light-cyan region).
A part of the disk surface could be blown in the wind, the flow of which
occurs only outside the acceleration region.
We call such a relatively dense flow ``the disk surface flow'' \citep{hac25kw}.
The velocity of the disk surface flow is slower than the nova wind.

Figure \ref{v1674_her_disk_config}(c) illustrates a successive phase.
The photosphere of nova winds expands faster leaving
the acceleration region behind, the position of which
is far inside the photosphere.
Because the density of the WD envelope is much lower 
than that of the accretion disk, 
the nova winds are expanding almost spherically but avoid the equatorial
region shaped by the dense disk as illustrated in Figure
\ref{v1674_her_disk_config}(b) and (c).

Before the wind blows, i.e., in the stage around
Figure \ref{v1674_her_disk_config}(a),
the total emission from the binary consists of the photospheric emission
(WD photosphere $+$ irradiated accretion disk photosphere $+$ companion star
photosphere),
although the companion star is not depicted in these plots. 
After the winds blow
(Figure \ref{v1674_her_disk_config}(b) and (c)),
the total emission from the binary consists of the photospheric emission
and free-free emission from the nova wind.

\subsection{Irradiated disk and companion star}
\label{irradiated_disk_companion}

In the very early phase of the outburst, the irradiated disk photosphere
and companion star photosphere dominantly contribute to the optical
luminosity of V1674 Her.  In our calculation, we simply assume that the
companion star fills its inner critical Roche lobe, as depicted
in Figure \ref{v1674_her_config}.  We assume different size and shape
of the accretion disk before and after the nova wind emerges from
the photosphere.

We first assume that the disk surface is cylindrically symmetric
around the WD.  Then, the outer size of the disk is defined by
\begin{equation}
R_{\rm disk}= \alpha R_{\rm RL,1},
\label{disk_radius_alpha}
\end{equation}
and the height of the disk at the outer edge is given by
\begin{equation}
H_{\rm disk}= \beta R_{\rm disk}.
\label{disk_height_beta}
\end{equation}
The surface height $z$ of the disk at the equatorial distance
$\varpi= \sqrt{x^2+y^2}$ from the center of the WD is assumed to be
\begin{equation}
z = \left({{\varpi} \over {R_{\rm disk}}}\right) H_{\rm disk}.
\label{disk_shape_winds}
\end{equation}

Before the nova wind blows, we adopt $\alpha = 0.85$ and $\beta=0.05$
as in Figure \ref{v1674_her_config}(a), where
$\alpha = 0.85$ is close to the tidal limit of the disk
\citep[see, e.g., ][]{mura24ki}. 
After the nova wind blows, we adopt $\alpha = 1.3$ and $\beta=0.05$
as in Figure \ref{v1674_her_config}(b).  This value is taken after
\citet{mura24ki} who found that the optically thick part of the disk
is extended up to $\alpha \sim 1.3$ during the nova wind phase in the
2021 outburst of U Sco.
We interpret that a large size disk over the tidal limit 
is a manifestation of the photosphere of the disk surface flow
(optically thick part).
It is reasonable that we assume axisymmetric flow for the disk surface
flow when the nova wind is axisymmetric, i.e., almost spherically symmetric
\citep[see, e.g., ][for more details]{hac25kw}.

We partition each photosphere of the WD, disk, and companion star into
many sectors (or patches) as in Figure \ref{v1674_her_config}.
Each small patch absorbs high energy photons from the WD.
The photospheric temperature of each patch increases.
We assume that the photosphere of each patch emits blackbody photons
at this increased temperature.  The $V$ light curve is calculated from
the summation of each patch with the standard $V$ filter response function.
See \citet{hac01kb} for the partitioning of each surface
and calculation method of irradiation.  
Note that, in Figure \ref{v1674_her_config}, gas is optically thin
outside the mesh surfaces
(i.e., photospheres of the disk, WD, and companion star).
In our irradiation calculation, we neglect absorption by optically thin
gas outside the mesh surfaces.

As introduced in Section \ref{not_disrupted_disk}, V1674 Her is
an IP \citep{pat22}.  The inner edge $R_{\rm in}$ of the disk 
is determined by the Alfven radius where the ram pressure of the inflowing
gas balances with the magnetic force of the WD
\citep[e.g., Equation (2) of ][$R_{\rm in}\sim 0.02 ~R_\sun$
for typical values]{qui24}.
We neglect this central hole of the accretion disk in our
luminosity calculation of the irradiated disk,
because the optical brightness of the accretion disk depends
on the surface area of the disk and a small innermost hole hardly 
contributes to the optical brightness of the disk. 

\begin{figure*}
\gridline{\fig{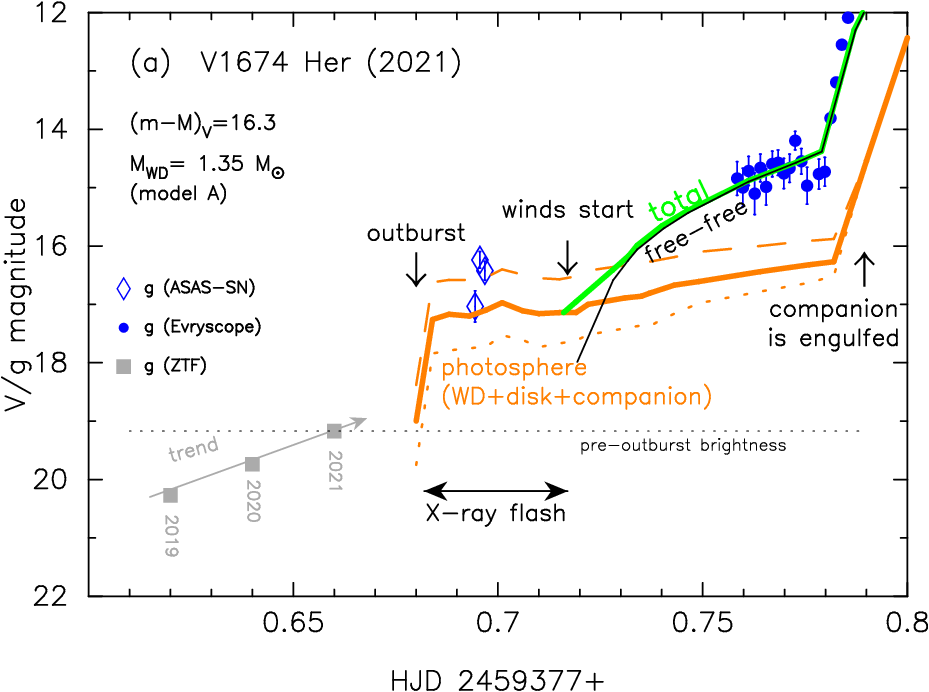}{0.65\textwidth}{}
          }
\gridline{\fig{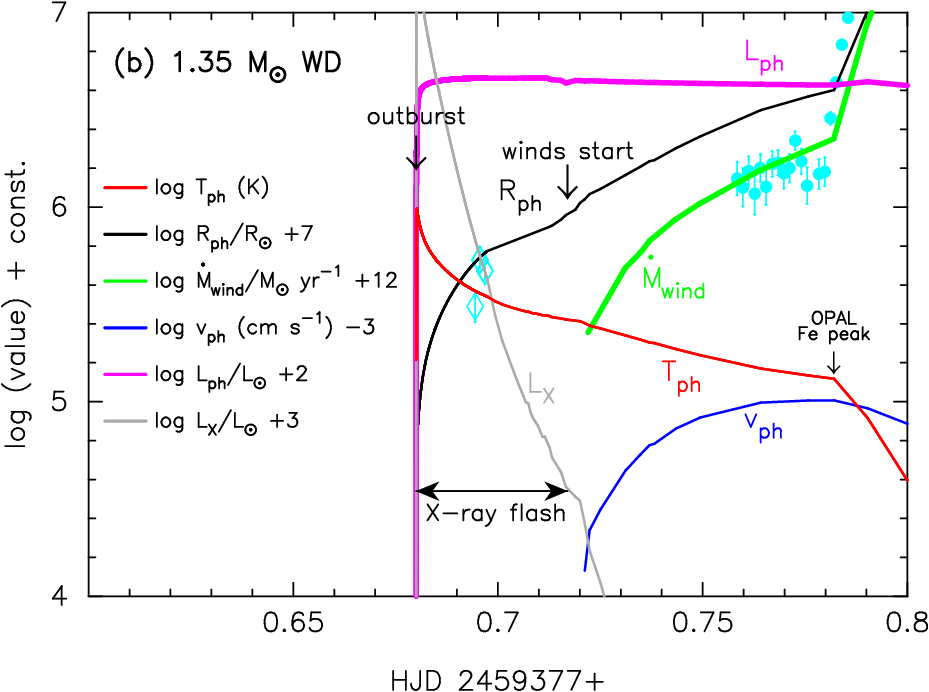}{0.65\textwidth}{}
          }
\caption{
(a) Comparison of our theoretical $V$ light curve with 
observational data in the first 0.12 days of the V1674 Her 2021 outburst. 
We assumed the origin of time ($t=0$) to be JD 2,459,377.68. 
The three orange lines (thick solid, dashed, and dotted) correspond to
the photospheric $V$ light curves of the binary with three different
inclination angles of $i=67$, 45, and $75\arcdeg$, respectively.
The black line shows the free-free emission model $V$ light curve and
the green line is the total $V$ luminosity of free-free (black line) and 
photospheric (thick solid orange line) luminosities.
Blue filled circles and blue open diamonds show
Evryscope and ASAS-SN $g$ magnitudes, respectively.
The data are all taken from \citet{qui24}.
The epoch when the companion star is engulfed by the WD photosphere
is indicated by the upward arrow for our WD model. 
A brightening trend of the Zwicky Transient Facility (ZTF)
$g$ magnitudes in the pre-outburst phase
is shown by the filled gray squares and the arrow penetrating them.
Note that its timescale is very different from that of abscissa,
in units of year. 
(b) Various photospheric properties of our $1.35 ~M_\sun$ WD model are
plotted during the same interval as in panel (a).  The data are all taken
from \citet{kat25hs}.}
\label{comparing_light_curve_v1674_her_early}
\end{figure*}

We include the effect of viscous heating by mass accretion
through the accretion disk \citep[see ][for details on this effect]{hac01kb}.
As the mass accretion increases, the temperatures of the disk surface
increases and, as a result, the disk brightens up.
\citet{qui24} showed that the pre-outburst brightness of V1674 Her
gradually increased up to $g=19.17$, 1.7 days before the outburst,
from the Zwicky Transient Facility (ZTF)
archive data (see the trend from 2019 to 2021
in Figure \ref{comparing_light_curve_v1674_her_early}(a)).
To reproduce $g=19.2$ before the outburst, we assume a mass accretion
rate to be as high as $\dot{M}_{\rm acc}= 2\times 10^{-7} ~M_\sun$ yr$^{-1}$.

\subsection{Composite light curves}
\label{temporal_variation_of_light_curve}

The $V$ magnitude luminosity from the WD photosphere $L_{V, \rm ph, WD}$
is calculated from the blackbody emission 
with standard $V$ response function. 
The $V$ luminosity from each photosphere is the summation of the photospheric
emission from the WD, irradiated accretion disk surface,
and the irradiated companion, which is expressed by
\begin{equation}
L_{V, \rm ph} = L_{V, \rm ph, WD} + L_{V, \rm ph, disk} + L_{V, \rm ph, comp}.
\label{photospheric_luminosity_v-band}
\end{equation}

After optically thick winds emerge from the WD photosphere,
free-free emission from the optically-thin ejecta outside the WD
photosphere begins to contribute to the $V$ luminosity.
We have calculated the composite $V$ light curve as
\begin{eqnarray}
L_{V, \rm total} &=& L_{V, \rm ff,wind} + L_{V, \rm ph, WD} \cr
  & & + L_{V, \rm ph, disk} + L_{V, \rm ph, comp},
\label{luminosity_summation_wd_disk_comp_v-band}
\end{eqnarray}
where $L_{V, \rm total}$ is the total $V$ luminosity of the V1674 Her system,
$L_{V, \rm ff,wind}$ the free-free emission $V$ luminosity. 

Figure \ref{comparing_light_curve_v1674_her_early}(a)
shows each component separately,
$L_{V, \rm ph}$ (Equation (\ref{photospheric_luminosity_v-band}):
thick orange line) and $L_{V, \rm ff,wind}$
(Equation (\ref{free-free_flux_v-band}): thin black line) as well as
the total emission $L_{V, \rm total}$
(Equation (\ref{luminosity_summation_wd_disk_comp_v-band}): thick green line).

The downward arrow labeled ``winds start'' indicates the epoch
when the nova wind emerges from the photosphere.  Before this epoch,
the irradiated disk and companion star are the main optical sources.
After winds emerges, free-free emission dominates as shown
by the green and black lines.

\section{X-ray flash phase light curve of V1674 Her}
\label{first_x-ray_flash}

Figure \ref{comparing_light_curve_v1674_her_early}(a) shows
the earliest optical light curve, the $V$ luminosity of which
is contributed mainly by the accretion disk and companion star photospheres
(thick orange line).
It is soon replaced by free-free emission from the optically-thin ejecta
(thin black line) after nova winds emerge from the WD photosphere.
Figure \ref{comparing_light_curve_v1674_her_early}(b) shows the $1.35 ~M_\sun$
WD properties, i.e., the photospheric luminosity $L_{\rm ph}$,
X-ray luminosity (0.3 - 1.0 keV) $L_{\rm X}$, temperature $T_{\rm ph}$,
radius $R_{\rm ph}$, velocity $v_{\rm ph}$, and wind mass-loss rate
$\dot{M}_{\rm wind}$.  The X-ray flux (gray line) sharply rises
just after the onset of thermonuclear runaway, $t_{\rm OB}=$HJD 2,459,377.68,
followed by a slower decline. We depict the X-ray flash phase
by the horizontal two-headed black arrow.
We see that the ASAS-SN $g$ data are obtained exactly during
the X-ray flash phase.  The $V$ luminosity is dominated by the irradiated
accretion disk and companion star (thick orange line).
In other words, the ASAS-SN $g$ data is the first optical detection 
of an X-ray flash.

In what follows, we describe details of our $V$ light curve model
step by step.

\subsection{Mass accretion rate on to the white dwarf}
\label{mass_accretion_rate}

\citet{qui24} showed that the pre-outburst brightness of V1674 Her
gradually increased up to $g=19.17$, until 1.7 days before the outburst,
from the ZTF archive data (see the trend from 2019 to 2021
in Figure \ref{comparing_light_curve_v1674_her_early}(a)).
The reason of the gradual brightening is unknown. If we attribute
this brightening to that of the accretion disk,
we need a mass accretion rate on to the WD as high as
$\dot{M}_{\rm acc}= 2\times 10^{-7} ~M_\sun$ yr$^{-1}$.
In other words, the disk with such a high mass accretion rate
is as bright as $V\sim 19$ owing to viscous heating.

This high mass accretion rate on to the WD through the accretion disk,
$\dot{M}_{\rm acc}= 2\times 10^{-7} ~M_\sun$ yr$^{-1}$,
is consistent with \citet{kat25hs}'s analysis.
V1674 Her shows a longer SSS phase than the expected
interval if no mass accretion occurs during the outburst.
They showed that the observed long SSS phase can be reproduced
if such a high mass accretion as 
$\dot{M}_{\rm acc}\sim 3\times 10^{-7} ~M_\sun$ yr$^{-1}$
had continued during the outburst
because much fresh fuel is supplied to hydrogen burning.
Otherwise, its duration is too short to be compatible with the observation.


\begin{figure*}
\epsscale{0.9}
\plotone{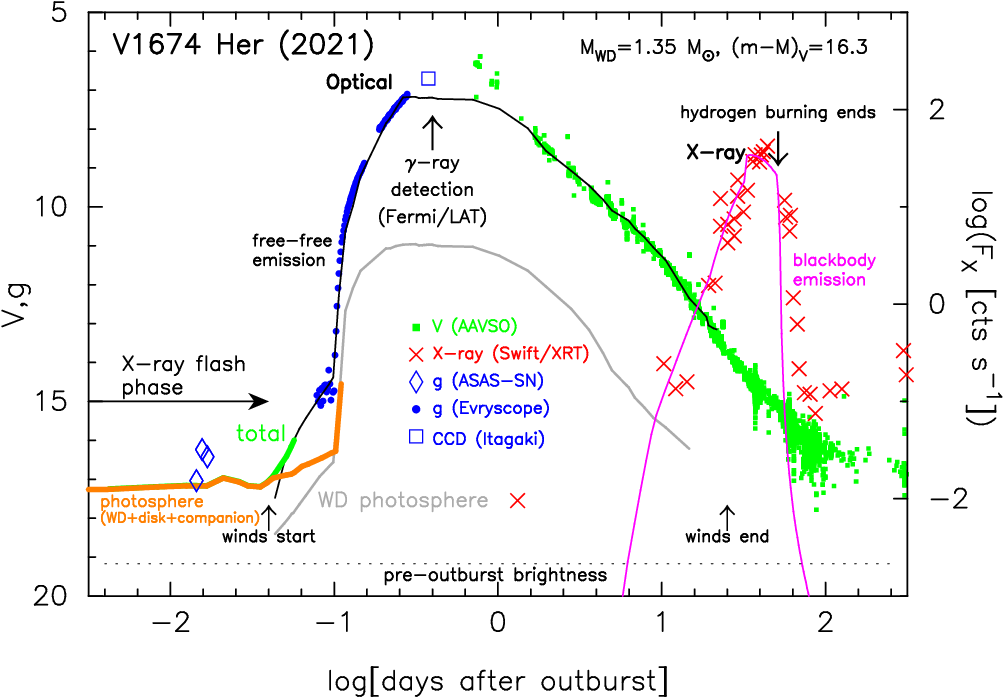}
\caption{
Same as Figure \ref{v1674_her_v_x_observation_only_logscale}, but we added
our light curve model (thick orange line) of the irradiated accretion disk
and companion star including the WD photosphere (light-gray line).
We also add the total flux (thick green line) of the free-free emission
(black line) plus photospheric blackbody emission (thick orange line).  
The $V$ luminosity of the WD photosphere is much smaller than
that of the free-free emission luminosity. 
\label{v1674_her_v_x_observation_only_logscale_no2}}
\end{figure*}

\subsection{More than 2 mag brightening-up after hydrogen burning}
\label{2mag_brightening-up}

Just after the nova outburst begins,
the photospheric brightness (thick solid orange line)
increases/jumps up from $g=19.2$ to $g=17.0$ mag.
The pre-outburst magnitude of $g=19.2$ is
owing to the viscous-heating accretion disk
whereas $g=17.0$ mag is from the irradiated disk and companion star.
A similar phenomenon of binary brightening-up of $\sim 2$ mag or more 
 was first definitely pointed out by \citet{hac25kdemarc}.
They proposed a demarcation criterion how to distinguish
if hydrogen-burning is on/off on a WD 
for millinova outbursts \citep{mro24ks}.
V1674 Her provides another example that hydrogen burning on
the WD makes $\Delta V$ (or $\Delta g)\sim 2$ mag increase
in the disk brightness.

\subsection{From photospheric emission to free-free emission}
\label{from_photospheric_to_free-free}

Figure \ref{v1674_her_v_x_observation_only_logscale_no2} depicts our
summary of light curves for the 2021 outburst of V1674 Her, which
combines  Figure \ref{v1674_her_v_x_observation_only_logscale} with
Figure \ref{comparing_light_curve_v1674_her_early}(a).
In the earliest phase before the wind starts, the light curve can be
explained by the emission from the irradiated photospheres of the disk
and companion star.  After the winds start, 
the free-free emission model $V$ light curve well follows
the Evryscope data (blue filled circles)
from the first data of $g=14.844$ (0.0784 day) until the last pre-maximum
data of $g=7.102$ (0.2791 day) over 7 magnitudes rise
including a break of slope at $g=14.3$ (0.1 day), where we assumed $V-g=0$.
This excellent agreement with the model and observation implies that,
in the rising phase from $g=14.9$ to $g=7.1$, 
there is no contribution of other sources such as a shocking power.

\begin{figure*}
\gridline{\fig{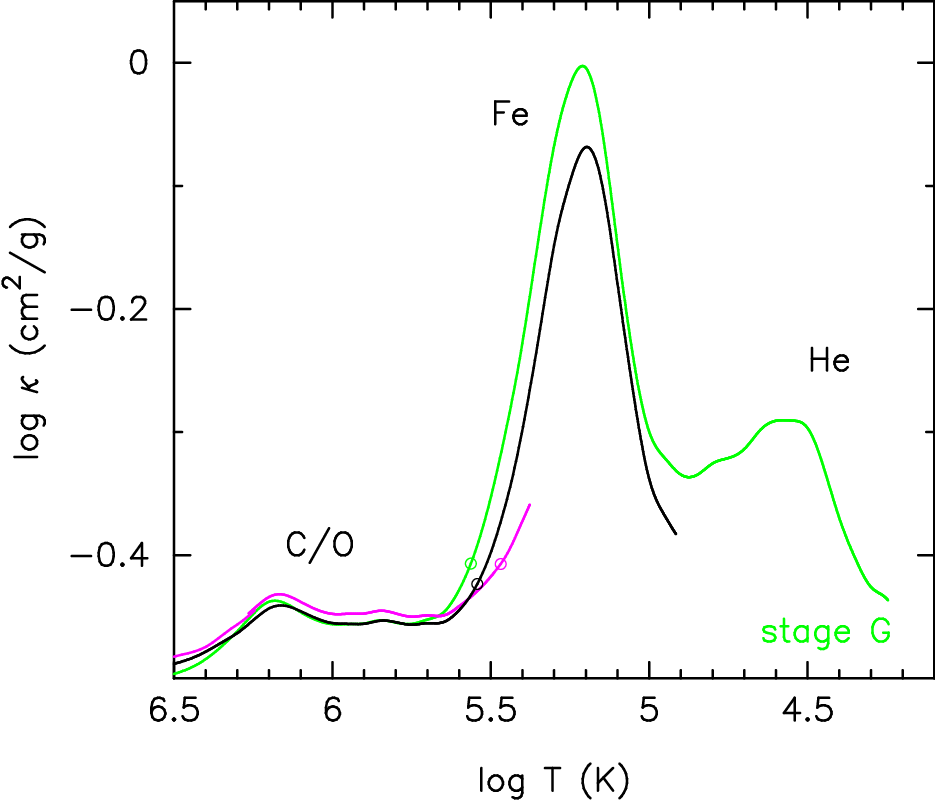}{0.45\textwidth}{(a) Run of the opacity}
          }
\gridline{\fig{f7b.eps}{0.35\textwidth}{(b) before Fe peak}
          \fig{f7c.eps}{0.35\textwidth}{(c) after Fe peak}
          }
\caption{
(a) Run of the radiative opacity against the temperature from inside
to outside in our $1.35 ~M_\sun$ WD for three stages.
The open circles on each line correspond to the critical points
of Parker type steady-state wind solutions \citep{kat94h}. 
Matter is sharply accelerated around this critical point. 
Magenta line: The stage of $\log T_{\rm ph} ({\rm K})=5.38$
(at $t=0.045$ day).  All the envelope still lies inside the Fe peak
($\log T\approx 5.2$) of the opacity.  
Black line: The stage of $\log T_{\rm ph} =4.92$ (at $t=0.11$ day). 
The envelope expands to include the Fe opacity peak and its photosphere
is outside the peak. 
Green line: stage G at the maximum expansion of the photosphere
(at $t=0.32$ day) in Figure \ref{hr}.  
(b) The diffusive luminosity $L_r$ and local Eddington luminosity
$L_{{\rm Edd}, r}$ are plotted against the radius $r$ 
for the stage of magenta line in panel (a).  
The open circle corresponds to the critical point of Parker
type supersonic wind.
(c) Same as panel (b), but for the stage of the black line in panel (a).
}
\label{opacity_luminosity_edd}
\end{figure*}

\section{Discussion}
\label{sec_discussion}

\subsection{Two-step rise in the free-fee emission light curve}
\label{two_step_rise}

Figure \ref{comparing_light_curve_v1674_her_early}(a) shows
that our free-free model $V$ light curve (black line) well follows
the temporal variation of the Evryscope $g$ 
light curve (blue filled circles) including before and after 
the break of slope on HJD 2,459,377.78.
This sudden rise in the $g$ magnitude is theoretically explained as follows:
The flux of free-free emission depends strongly on the wind
mass-loss rate, i.e., $L_{V, \rm ff,wind}\propto (\dot{M}_{\rm wind})^2$ 
from Equation (\ref{free-free_flux_v-band}),
which suddenly increases at the break.

This sudden change of the wind mass loss rate is closely related to
the interior structure of the nova envelope.
Figure \ref{opacity_luminosity_edd}(a) plots the opacity $\kappa$
distributions \citep[OPAL, ][]{igl96} in the envelope against
the temperature for three stages, slightly before (magenta
line) and after (black line) the epoch of the break,
and at the maximum expansion of the WD photosphere (green line), stage G. 
There are small and large opacity peaks, which correspond to
the ionization zone of C/O, Fe, and He, from inside to outside (from high
temperature to low temperature).  The right edge of each line corresponds
to the photosphere.  In an early stage (the magenta line in Figure
\ref{opacity_luminosity_edd}(a)) before the $g$ magnitude break, 
the photospheric temperature is so high and the photosphere is inside
the strong Fe peak. In a later stage (the black line in Figure
\ref{opacity_luminosity_edd}(a)) after the $g$ magnitude break,
the envelope has expanded so that the photospheric temperature
decreases down outside the Fe peak.
This difference causes weak (before) and strong (after) 
acceleration of winds.

We plot the distributions of the photon luminosity $L_r$ and
local Eddington luminosity defined by
\begin{equation}
L_{{\rm Edd}, r} \equiv {4\pi cG{M_{\rm WD}} \over\kappa},
\label{equation_local_Edd}
\end{equation}
where $L_r$ the photon diffusive luminosity at the radius $r$
from the WD center, $L_{{\rm Edd}, r}$ is the local Eddington luminosity
at the radius $r$, and the opacity $\kappa$ is a function of
the density and temperature, $c$ the speed of light, and $G$ the
gravitational constant.
Figure \ref{opacity_luminosity_edd}(b) depicts the $L_r$ (black line) and 
$L_{{\rm Edd}, r}$ (red line) distributions
slightly before the break, corresponding
to the magenta line in Figure \ref{opacity_luminosity_edd}(a), while
Figure \ref{opacity_luminosity_edd}(c) plots the distributions
slightly after the break, corresponding to the black line in Figure
\ref{opacity_luminosity_edd}(a).

The optically thick winds are accelerated by radiative pressure
gradient/difference deep inside the photosphere \citep{fri66, kat94h}. 
The radiation pressure gradient/difference is simply written by 
$\Delta P_{\rm rad} \propto (L_r - L_{{\rm Edd}, r})$
for a local super-Eddington region of $L_r - L_{{\rm Edd}, r} > 0$.
Here, $P_{\rm rad}$ is the radiation pressure. 
We see in Figure \ref{opacity_luminosity_edd}(b) and (c) 
that the difference of $L_r - L_{{\rm Edd}, r}$ sharply increases 
as the envelope expands and reaches maximum at the break. 
The break occurs just at the photospheric temperature of  
$\log T_{\rm ph}~({\rm K}) \sim 5.2$, as shown
by the red line in Figure \ref{comparing_light_curve_v1674_her_early}(b).
This corresponds to the temperature at the Fe peak of the opacity $\kappa$
as in Figure \ref{opacity_luminosity_edd}(a).
After the break (at $\log T_{\rm ph}~({\rm K}) \sim 5.2$),
the difference, $L_r - L_{{\rm Edd}, r}$, gradually increases 
always keeping the Fe peak of $\kappa$ inside the photosphere. 
Comparing the difference between before 
(Figure \ref{opacity_luminosity_edd}(b)) and after
(Figure \ref{opacity_luminosity_edd}(c)) the break, the acceleration
force of the wind simply increases by a factor of several. 
This is the reason of sudden increase of 
the wind mass loss rate and thus, sudden increase of the rising slope
in the free-free emission light curve at/near 
$\log T_{\rm ph}~({\rm K}) \sim 5.2$. 

\citet{qui24} noted that the early Evryscope $g$ data indicate a 
shallow dip between the slow and fast rises 
which does not appear in \citet{kat25hs}'s model as shown 
in Figure \ref{comparing_light_curve_v1674_her_early}(a).
If this optical dip is real, it would suggest 
that a rapid structure change near the photosphere 
can not be followed by the numerical method of Kato et al. 
who adopted a steady-state wind approximation for each time step.


\begin{figure}
\epsscale{0.9}
\plotone{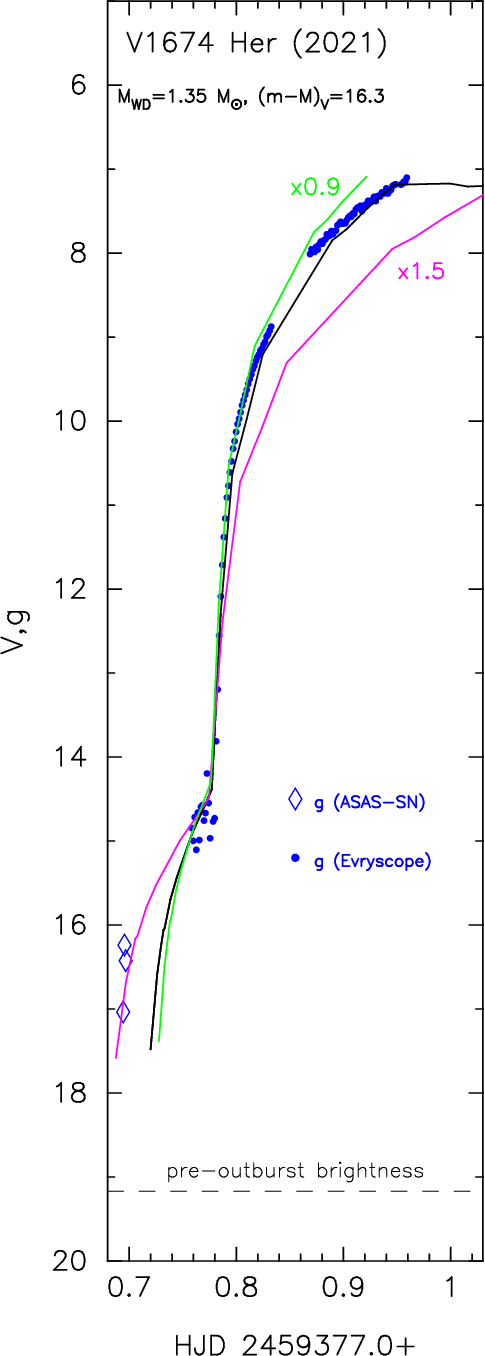}
\caption{
The $g$ band light curves (blue symbols) and our free-free emission model
$V$ light curve (black line) in the very early phase of the V1674 Her
outburst.  The observational data and free-free emission model $V$
light curve (black line) are the same as those in Figure 
\ref{v1674_her_v_x_observation_only_logscale_no2}.
We add other two model light curves (magenta and green lines),
the timescales of which is 1.5 and 0.9 times longer than that of
the original model light curve (black line).  We nail down each
break of the light curves on HJD 2,459,377.78.
The wind phase of magenta line starts just at the first ASAS-SN detection.
\label{v1674_her_v_rising_observation_linear_no2}}
\end{figure}

\subsection{Inclination angle of the binary}
\label{binary_inclination_angle}

We assumed the inclination angle $i=67\arcdeg$ in our calculation of the
light curve. The resultant brightness of $V\sim 17$ is consistent with the
first ASAS-SN observation of $g=17.0$ at $t=0.0144$ day (HJD 2,459,377.6944)
as shown in Figure \ref{comparing_light_curve_v1674_her_early}(a). 
Slightly ($\sim 105$ and 210 s) later, the $g$ magnitude brightens up to
$g= 16.24$ ($t=0.0156$ day) and $g=16.42$ ($t=0.0168$ day).

This short interval is consistent with the timescale of spin period of the WD,
$P_{\rm spin}=501$ s, as discussed by \citet{qui24}.
However, the WD photosphere has already expanded up to
$R_{\rm ph}\sim 0.05 ~R_\sun$ at $t=0.0144$ day ($=21$ min) and
has completely engulfed the magnetic funnels of the innermost region of
the accretion disk if the inner edge of the disk
is as small as $R_{\rm in}\sim 0.02 ~R_\sun$ 
\citep[a typical value of Equation (2) in ][]{qui24}.  
So, it is unlikely that this very short timescale variation is related to
the spin of the WD unless $R_{\rm in}$ is as large as 
$\gtrsim 0.06 ~R_\sun$.  Such a larger hole in the disk could be possible
for some extreme cases for V1674 Her.  Substituting the WD mass
of $M_{\rm WD}=1.35 ~M_\sun$, mass accretion rate of
$\dot{M}_{\rm acc}=2\times 10^{-7} ~M_\sun$ yr$^{-1}$,
and magnetic field of $B=10^7$ G into Equation (2) of \citet{qui24},
we obtain $R_{\rm in}\approx 0.056 ~R_\sun$.  Or substituting
$1.35 ~M_\sun$, $\dot{M}_{\rm acc}=1\times 10^{-11} ~M_\sun$ yr$^{-1}$,
and $B=10^5$ G, we have $R_{\rm in}\approx 0.069 ~R_\sun$.
If it is the case, this short timescale variation could be explained
by some geometric effect of the disk and magnetic funnels,  
as suggested by \citet{qui24}.  

When $R_{\rm in}\lesssim 0.05 ~R_\sun$,
we do not know the origin of this short timescale variability.
Because the first data has relatively large statistic error
of $\sigma_g=0.3$ mag, we searched other model light curves that fit
the second and third data by changing the inclination angle of the binary.
As shown in Figure \ref{comparing_light_curve_v1674_her_early}(a),
a smaller inclination angle of $i=45\arcdeg$ (dashed orange line) makes
magnitude 0.5 mag brighter than the $i=67\arcdeg$ case (thick solid
orange line) that is broadly consistent with the ASAS-SN $g=16.4$ magnitudes.
Assuming that the second and third data are securer than the first one,
we prefer to the inclination angle of $i=45\arcdeg$ rather than $i=67\arcdeg$.

\subsection{Timescale of the optical rise}
\label{timescale_optical_rise}

The earliest ASAS-SN $g$ data were obtained in the X-ray flash phase of
V1674 Her as shown in Figure \ref{comparing_light_curve_v1674_her_early}. 
Here, we discuss another possibility that the ASAS-SN data were explained
by the free-free emission light curve, not by the irradiation effect of
a disk in the X-ray flash phase. 

The most important parameters for the light curve fitting is the WD mass.
\citet{kat25hs} presented only two light curves of $1.35 ~M_\sun$ WD
with the mass accretion rate of $\dot{M}_{\rm acc}=1\times 10^{-11} ~M_\sun$
yr$^{-1}$ (model A) and $5\times 10^{-10}~M_\sun$ yr$^{-1}$ (model B).
Although they claimed they did not intend to make a fine tuning model,
their model with the lower mass accretion rate 
happens to show an excellent agreement with the V1674 Her light curve data,  
both in the fast rising phase and decay phase. 
If they adopt a massive (less massive) WD, the light curve will be 
faster (slower) and does not match the observation.  
Therefore, we conclude that $1.35 ~M_\sun$ is a quite good WD mass
for V1674 Her. 

The second important parameter is the mass accretion rate.
If we adopt a larger (smaller) mass accretion rate, the rising phase
becomes slower (faster) as shown by \citet{kat25hs}.
Their model B (the higher mass accretion rate 
of $5\times 10^{-10}~M_\sun$ yr$^{-1}$) shows a much slower rise
than that of model A (our original model in the present paper). 
The earliest ASAS-SN data correspond to the phase 
after the winds started, and were explained by the free-free emission 
model light curve (not by the irradiation of the disk). 
This model B, however, shows much slower rise from $g=14$ toward the peak
as shown in Figure 7 of \citet{kat25hs}, and is inconsistent with
the observation.  We safely exclude this model B.

The optical rising timescale of model B is about 2.5 times slower than that
of model A.   In Figure \ref{v1674_her_v_rising_observation_linear_no2},
we plot a 1.5 times longer case (magenta line) than that of model A.
Here, we nail down the break in the rising slope of each light curve
at HJD 2,459,377.78.  The magenta line of free-free emission model $V$
light curve starts from the earliest ASAS-SN data, but the final rise
is clearly different from the observation.  Therefore, we safely reject
the possibility that the ASAS-SN data enter the WD wind phase of the nova.
We conclude that the earliest ASAS-SN $g$ data fall in the X-ray
flash phase.

\subsection{Outburst day}
\label{ourburst_day}

We adopt the outburst day of $t=0=t_{\rm OB}=$ HJD 2,459,377.68 after
\citet{kat25hs}.  If we could shift the outburst day slightly later
to $t_{\rm OB}=$ HJD 2,459,377.69, the earliest three ASAS-SN data 
were observed just on this outburst day (Figure 
\ref{comparing_light_curve_v1674_her_early}(a)).
In this case, a brightening-up of ASAS-SN data from $g=17$ to 16.2 mag
could occur within 0.0012 day ($=105$ sec). 

Then, the timescale of the light curve is 0.9 times compressed 
compraed with our original model. We plot such a model (green line)
in Figure \ref{v1674_her_v_rising_observation_linear_no2}. 
The difference between the black line and the green line should be detected
because of plenty of observational data near $g=8$--7.
Thus, we safely exclude the case of $t=0=t_{\rm OB}=$ HJD 2,459,377.69
as the outburst day.

\section{Conclusions}
\label{sec_conclusion}

Our results are summarized as follows: 
\begin{enumerate}
\item We present a light curve model for the earliest phase
of the V1674 Her outburst, based on the binary model consisting of
a $1.35~M_\sun$ WD, an irradiated accretion disk and companion star.
For the nova explosion model, we used the fully self-consistent
nova outburst model of a $1.35~M_\sun$ WD with a mass accretion rate of
$1\times 10^{-11}~M_\sun$~yr$^{-1}$ \citep{kat25hs}.

\item The first detected optical data (ASAS-SN $g=17.0$) correspond to
$t=0.34$ hr (0.0144 day) after the onset of thermonuclear runaway,
which fall in the X-ray flash phase. 
The brightness of $g=17$ mag is consistent with that of an
irradiated accretion disk. 
This ASAS-SN $g$ data clearly shows the X-ray flash 
phase of a classical nova first detected with an optical band. 

\item The $g$ brightness increases/jumps up from the pre-outburst
brightness of $g=19.2$ to $g=17.0$ mag in the X-ray flash phase.
This $\Delta g\sim 2.2$ mag increase is owing to the irradiation effects
of the accretion disk.
A similar 2.2 mag jumping-up/increase by irradiation effects
was first pointed out in millinova outbursts by \citet{hac25kdemarc}.
V1674 Her adds another example to this law that hydrogen ignition on
a WD makes more than 2.2 mag increase in the disk optical brightness.\\

\item Optically thick winds emerge from the WD photosphere at $t=0.96$ hr
(0.04 day) after the outburst.  As the wind mass-loss rate increases, 
the dominant source of optical emission changes,
after $t=1.1$ hr ($=$0.046 day), from photospheric emission
of the WD, disk, and companion star to free-free emission of
optically-thin plasma just outside the WD photosphere.
Our free-free emission model $V$ light curve reproduces the Evryscope
$g$ light curve from $g=14.8$ (at 0.0784 day) to $g=7.1$ (at 0.2791 day)
including a break of light curve slope at $g=14.3$ (on 0.1 day),
where we assume $V-g=0$.   

\item The break in the slope of our free-free emission model light curve
is caused by a large increase in the acceleration force due to
the strong Fe peak of OPAL opacity.
This is the first manifestation of the Fe peak of opacity in the
classical nova light curves.  

\item There is no indication of strong shocking power in the optical 
rise between $g=14.8$ and $g=7.1$.
\end{enumerate}

\begin{acknowledgments}
We acknowledge with thanks the variable star observations (V1674 Her)
from the AAVSO International Database contributed by observers worldwide
and used in this research.
We are also grateful to the anonymous referee for useful comments
that improved the manuscript.
\end{acknowledgments}

\vspace{5mm}
\facilities{Swift(XRT), AAVSO}




\begin{thebibliography}{}




\bibitem[Y. Bhargava et al. (2024)]{bha24}
Bhargava, Y., Dewangan, G. C., Anupama, G.,C., et al.
2024, \mnras, 528, 28, \doi{10.1093/mnras/stad3870}







\bibitem[J. J. Drake et al. (2021)]{dra21}
Drake, J. J., Ness, J.-U., Page, K. L., et al. 2021, \apjl, 922, L42,
\doi{10.3847/2041-8213/ac34fd}



\bibitem[P. A. Evans et al. (2009)]{eva09}
Evans, P. A., Beardmore, A. P., Page, K. L., et al.  2009, \mnras, 397, 1177,
\doi{10.1111/j.1365-2966.2009.14913.x}

\bibitem[M. Friedjung (1966)]{fri66}
Friedjung, M. 1966, \mnras, 132, 317, \doi{10.1093/mnras/132.2.317}




\bibitem[G. R. Habtie et al. (2024)]{hab24}
Habtie, G. R. Das, R., Pandey, R., Ashok, N.M., \& Dubovsky, P.A. 2024,
\mnras, 527, 1405, \doi{10.1093/mnras/stad3295}

\bibitem[Hachisu \& Kato (2001)]{hac01kb}
Hachisu, I., \& Kato, M. 2001, \apj, 558, 323, \doi{10.1086/321601}




\bibitem[I. Hachisu \& M. Kato (2006)]{hac06kb}
Hachisu, I., \& Kato, M. 2006, \apjs, 167, 59, \doi{10.1086/508063}



%






\bibitem[I. Hachisu \& M. Kato (2022)]{hac22k}
Hachisu, I., \& Kato, M. 2022, \apj, 939, 1, \doi{10.3847/1538-4357/ac9475}



\bibitem[I. Hachisu \& M. Kato (2025)]{hac25kdemarc}
Hachisu, I., \& Kato, M. 2025, \apj, 983, 145, \doi{10.3847/1538-4357/adc107}




\bibitem[I. Hachisu et al. (2025)]{hac25kw}
Hachisu, I., Kato, M., \& Walter, F. M. 2025, \apj, 980, 142, 
\doi{10.3847/1538-4357/adae08}


\bibitem[I. Hachisu et al. (2020)]{hac20skhs}
Hachisu, I., Saio, H., Kato, M., Henze, M., \& Shafter, A. W. 2020,
\apj, 902, 91, \doi{10.3847/1538-4357/abb5fa}



\bibitem[C.A. Iglesias \& F.J. Rogers (1996)]{igl96}
Iglesias, C. A., \& Rogers, F. J. 1996, \apj, 464, 943, \doi{10.1086/177381}


\bibitem[M. Kato \& I. Hachisu (1994)]{kat94h}
Kato, M., \& Hachisu, I., 1994, \apj, 437, 802, \doi{10.1086/175041}



\bibitem[M. Kato et al. (2025)]{kat25hs}
Kato, M., Hachisu, I., \& Saio, H. 2025, \apj, in press (arXiv:2506.04615)
\doi{10.48550/arXiv.2506.04615}




\bibitem[M. Kato et al. (2022a)]{kat22sha}
Kato, M., Saio, H., \& Hachisu, I. 2022a, \pasj, 74, 1005,
\doi{10.1093/pasj/psac051}

\bibitem[M. Kato et al. (2022b)]{kat22shapjl}
Kato, M., Saio, H, \& Hachisu, I. 2022b, \apjl, 935, L15,
\doi{10.3847/2041-8213/ac85cl}

\bibitem[M. Kato et al. (2022c)]{kat22shc}
Kato, M., Saio, H, \& Hachisu, I. 2022c, Research notes of the AAS, 6, 258,
\doi{10.3847/2515-5172/aca8af}


\bibitem[O. K{\"o}nig et al. (2022)]{kon22wa}
K{\"o}nig, O., Wilms, J., Arcodia, R., et al. 2022, \nat, 605, 248,
\doi{10.1038/s41586-022-04635-y}


\bibitem[C.-C. Lin et al. (2022)]{lin22}
Lin, L, C.-C., Fan, J.-L., Hu, C.-P., Tanaka, J., \& Li, K.-L. 2022,
\mnras, 517, L97, \doi{10.1093/mnrasl/slac117}






\bibitem[P. Mr\'oz et al. (2024)]{mro24ks}
Mr\'oz, P., Kr\'ol, K., Szegedi, K., et al. 2024, \apjl, 977, L37,
\doi{10.3847/2041-8213/ad969b}


\bibitem[Muraoka et al. (2024)]{mura24ki}
Muraoka, K., Kojiguchi, N., Ito, J., et al. 2024, \pasj, 76, 293,
\doi{10.1093/pasj/psae010}

\bibitem[M. Orio et al. (2022)]{ori22}
Orio, M., Gendreau, K., Giese, M., et al.
2022, \apj, 932, 45, \doi{10.103847/1538-4357/ac63be}

\bibitem[J. Patterson et al. (2022)]{pat22}
Patterson, J., Enenstein, J, de Miguel, E.,  et al. 2022, \apjl, 940, L56,
\doi{10.3847/2041-8213/ac9ebe}


\bibitem[R. M. Quimby et al. (2024)]{qui24}
Quimby, R. M., Metzger, B. D., 
Shen, K.J., et al. 2024, \apj, 977, 17, \doi{10.3847/1538-4357/ad887f}









\bibitem[K. V. Sokolovsky et al. (2023)]{sok23}
Sokolovsky, K. V., Johnson, T.J., Buson, S., et al. 2023, \mnras, 521,5453,
\doi{10.1093/mnras/stad887}





\bibitem[C. E. Woodward et al. (2021)]{woo21}
 Woodward, C. E., Banerjee D. P.K., Geballe, T.R. et al. 2021,
\apjl, 922, L10, \doi{10.3847/2041-8213/ac3518}



\end{thebibliography}
\end{document}